\documentclass[prc,showpacs]{revtex4}
\usepackage{graphicx}                                                          
\begin{document}
\title{
\hfill{\small {\bf MKPH-T-03-15}}\\
{\bf Final state interaction in the reaction $NN\to\eta NN$ }
}
\author{A.\,Fix\footnote{On leave of absence from Tomsk Polytechnic 
University, 634034 Tomsk, Russia} 
and H.\,Arenh\"ovel}
\affiliation{Institut f\"ur Kernphysik,
Johannes Gutenberg-Universit\"at Mainz, D-55099 Mainz, Germany}
\date{\today}
  
\begin{abstract}
The production of $\eta$ mesons in $NN$ collisions in the near threshold
region is analyzed with special emphasis on the role of final state
interaction which is calculated using three-body scattering theory. The
three-body aspects are shown to be very important in interpreting 
experimental results for the total cross section of $pp\to\eta pp$. 
Furthermore, energy spectra and angular distributions of the
emitted particles are calculated and compared to experimental data.
\end{abstract}

\maketitle


\section{Introduction}
The production of $\eta$ mesons in two-nucleon collisions depends strongly
on the final state interaction (FSI) as has been stressed in 
previous work~\cite{Wilkin01,Gedal98,Pena01,Nak02,Baru03,Del03}. 
Notwithstanding this fact, the interaction between the final particles
was treated in quite an approximate way. Only the $NN$ interaction has
been considered in a more precise manner. The interaction of the
$\eta$ meson either was not considered at all or simulated by means of
a Watson enhancement factor. In view of this apparent lack of a
correct treatment of FSI we want to study this reaction by using a
rigorous three-body model.  

The properties of the $\eta NN$ system, containing three free
particles in the asymptotic region, were investigated in \cite{FiAr02}
where FSI effects in $\eta-d$ reactions as well as in $\eta$
photoproduction on a deuteron were studied. As was shown in detail,
the three-body aspects of FSI are very important, at least near the
threshold region where $S$ waves give by far the largest
contribution to the production process. It is obvious that the
three-body $S$ wave part is most sensitive to FSI in view of the
dominance of the $S$ wave interaction in the $NN$ and $\eta N$
subsystems at low energies. Furthermore, it is intuitively clear that
primarily the central region of the system, where $S$ waves dominate, is
distorted by multiple scattering events. As a consequence, close to
threshold the $\eta NN$ interaction strongly enhances the $\gamma
d\to\eta np$ cross section. At higher energies, where the large spatial
extension of the deuteron comes into play, an essential fraction of
higher partial waves contributes increasingly to the $\eta$ production
mechanism. This rapidly reduces the relative $S$ wave contribution to
the reaction amplitude, so that the role of FSI diminishes
considerably with increasing energy. As a result, already in the
region of c.m.\ excess energies above 100 MeV the behavior of the
cross section is well explained in the plane wave approximation for
the final state~\cite{FiAr97}. 

Turning now to the process $NN\to \eta NN$, firstly we would like to note 
that in this case the influence of FSI must be even more important. Here the
characteristic radius $R$ of the reaction, determined by $R\sim 1/q$ with 
$q$ as typical momentum transfer, is only about 0.3~fm. This value is much 
smaller than in photoproduction, where not too close to the threshold
the characteristic range is given by the deuteron radius $r_d$, i.e.,
$R \sim r_d \approx$ 2~fm. Therefore, one may expect that for energies
for which $pR < 1$ is valid, with $p$ denoting the maximal c.m.\
momentum of the final particles, the final state will be dominated by
$S$ waves, for which the role of the $\eta NN$ interaction is
crucial. Indeed, as will be shown below, in the region from threshold
up to an excess energy $Q = 60$ MeV, the contribution of higher
partial waves in the final state is of much minor importance compared
to the $S$ wave contribution. In this connection, the reaction $NN\to
\eta NN$ seems in general to be more suited for studying the
interaction in the $\eta NN$ system than $\eta$ photoproduction, where
FSI has a smaller influence. This advantage is however strongly reduced by a
poor knowledge of the $\eta$ production mechanism in $NN$ collisions. 

In the present paper we investigate the reactions $np\to\eta np$ and
$pp\to\eta pp$ in the region of excess energy up to $Q = 60$ MeV which
is covered in recent experiments as well as in theoretical studies 
(see e.g.\ \cite{Mosk03} and references therein). The two questions 
which we address here are:

(i) To what extent are the three-body aspects essential for a quantitative
description of FSI? As was already stressed in~\cite{FiAr03}, the
importance of a three-body treatment of FSI is dictated by the
proximity of a virtual pole in the $\eta NN$ amplitude to the physical
region~\cite{FiAr00}. Because of this property a perturbative approach
to the $\eta NN$ dynamics, where only the interactions in the two-body
$NN$ and $\eta N$ subsystems are taken into account, is of little
use. In the present study we would like to demonstrate this fact for
$\eta$ production in $NN$ collisions. 

(ii) What is the role of higher $NN$ partial waves? According to
Nakayama \cite{Nak03}, the puzzling second bump in the $pp$ invariant
mass distribution of $pp\to\eta pp$ measured at $Q = 41$ MeV
in~\cite{TOF} could be understood as arising mostly from a
contribution of the $^3P_0$ state of the final nucleon pair. However,
the size of $P$ wave contribution appears
to be comparable with that of the $^1S_0$ one (see
e.g.~Fig.~4 in~\cite{Nak03}). Such a large $P$ wave effect in the
region where the $NN$ c.m.\ kinetic energy is limited to $T_{NN} < Q =
41$~MeV (and in fact the maximum of the distribution of $T_{NN}$ is
located at even a considerably lower value because of the preference
of higher $\eta$ momenta) is quite surprising. 
Therefore, we have investigated again the
influence of $S$, $P$, and $D$ partial $NN$ waves in both 
isospin states using the Paris $NN$ potential. Another point, closely
connected to this, is the role of the short range behaviour of the
$NN$ interaction which is enhanced by quite large momentum transfers
accompanying the $\eta$ production. As will be shown below, this leads
in particular to an important role of the tensor force in the final
$np$ triplet state for $np\to\eta np$ because the $^3D_1$ state,
excited in the $\eta$ production process, couples to the $^3S_1$
partial wave during subsequent $NN$ rescattering. 

First we review briefly in the next section the most essential formal
ingredients on which our calculation is based. For the driving $NN\to
NN^*$ transition potential we take a meson exchange model including
$\pi$, $\rho$, and $\eta$ exchanges. The two-body $\eta N$ and $NN$
interactions are parametrized in terms of separable $t$ matrices. The
dominant $S$ wave part of the $\eta NN$ wave function is calculated
using three-body scattering theory. In Sect.~\ref{sect2} our results
are compared with experimental data and in Sect.~\ref{sect4} the most
important conclusions are summarized. In an appendix we list detailed
expressions for the $NN\to NN^*$ amplitude. 


\section{Formal ingredients}\label{sect1}

In the center of mass system the reaction to be studied reads in detail
\begin{equation} \label{10}
N(E,\vec{p}\,)+N(E,-\vec{p}\,)\to N(E_1,\vec{p}_1)+N(E_2,\vec{p}_2)
+\eta(\omega_\eta,\vec{q}\,)\,,
\end{equation}
where the energies and momenta of the participating particles are
explicitly defined for later purpose. The reaction is governed by the
transition matrix element 
\begin{equation} \label{15}
M_{TM_T,S'M_S',SM_S}=\langle \vec{q};\vec{p}_1,\vec{p}_2;S'M_{S'};
TM_T|\,\hat{M}\,|SM_{S};TM_T,\vec {p}\, \rangle\,, 
\end{equation}
where initial and final states are characterized by the total spin
$S\,(S')$ and isospin $T$ of the nucleon pair. The calculation is done
in the region of c.m.\ excess energy $Q$ up to 60 MeV. Thus the use of
nonrelativistic expressions for the energies of the final particles is
justified, whereas for the initial state relativistic expressions are
taken. 

The c.m.\ cross section for $pp\to\eta pp$ is then given by
\begin{equation} \label{20}
\frac{d\sigma}{dq\,d\Omega_q\, d\Omega_p}
=\frac{1}{(2\pi)^5}\,\frac{M_N^3\ q^2p_r}{4Wm_\eta p}\
\frac{1}{4}\sum\limits_{S'M_S',SM_S}\!|M_{11,S'M_S',SM_S}|^2\,,
\end{equation}
where the relative momentum of the two final nucleons
$\vec{p}_r=(\vec{p}_1-\vec{p}_2)/2$ is introduced. The factor
$\frac{1}{4}$ results from the average over initial spin states. The
corresponding $np\to\eta np$ cross section is obtained from (\ref{20})
by the substitution 
\begin{equation} \label{21}
M_{11,S'M_S',SM_S}\to \frac{1}{2}\sum\limits_{T=0,1}M_{T0,S'M_S',SM_S}\,.
\end{equation}
For the $\eta$ production mechanism we have adopted the usual model
where first an $N^*$ resonance is excited on one of the nucleons which
subsequently decays into the $\eta N$ channel. The submatrix
$M_{NN^*}$ associated with the transition $NN\to NN^*$ is expressed as
a linear combination of eight pseudoscalar operators $\Sigma_j$
($j=1,8$) built from the vectors $\vec{p}$, $\vec{p}\,'$, and spin
operators $\vec{\sigma}_1$ and $\vec{\sigma}_2$ 
\begin{equation}\label{25}
M_{NN^*}(\vec{p},\vec{p}\,')=\sum\limits_{j=1}^{8}M_j\,
\Sigma_j(\vec{p},\vec{p}\,',\vec{\sigma}_1, \vec{\sigma}_2)\,,
\end{equation}
where $\vec{p}\,'$ denotes the relative momentum of the $NN^*$ system. 
The amplitudes $M_j$ are functions of only $p=|\vec p\,|$, $p'=|\vec
p^{\,\prime}|$ and $\vec{p}\cdot\vec{p}\,'$, and their specific 
form depends on the model for the $\eta$ production mechanism. 
For the operators $\Sigma_j$ we have chosen the following forms
\begin{eqnarray}\label{30}
\Sigma_{1(2)}&=&\vec{\sigma}_1\cdot\vec{p}^{\,\,(\prime)}\,, \nonumber\\
\Sigma_{3(4)}&=&\vec{\sigma}_2\cdot\vec{p}^{\,\,(\prime)}\,, \nonumber\\ 
\Sigma_{5(6)}&=&\frac{i}{2}(\vec{\sigma}_1\times\vec{\sigma}_2)
\cdot\vec{p}^{\,\,(\prime)}\,,\nonumber\\  
\Sigma_{7(8)}&=&i\sqrt{5}\left\{(\vec{\sigma}_1\otimes\vec{\sigma}_2)^{[2]} 
\otimes(\vec{p}^{\,\,(\prime)}\otimes(\vec{p}\times\vec{p}\,'))^{[2]}
\right\}^{[0]}\,.
\end{eqnarray}
The latter two operators are tensor operators in two-nucleon spin
space describing a spin transfer of 2 in the transition to the $NN^*$
state. For the impulse approximation (IA), where FSI is ignored, the
amplitude $M_{TM_T,S'M_S',SM_S}$ (\ref{15}) is given by 
\begin{eqnarray}\label{35}
M^{(\mathrm{IA})}_{TM_T,S'M_S',SM_S}&=&\langle S'M_{S'};TM_T| 
\Big[\Big(M_{NN^*}(\vec{p},\vec{p}_2)-(-1)^{S+T}M_{NN^*}
(-\vec{p},\vec{p}_2)\Big)F_{N^*}(\vec{q},\vec{p}_1)\nonumber\\
&&-(-1)^{S'+T}\Big(M_{NN^*}(\vec{p},\vec{p}_1)
-(-1)^{S+T}M_{NN^*}(-\vec{p},\vec{p}_1)\Big)F_{N^*}(\vec{q},\vec{p}_2)\Big]
|SM_{S};TM_T \rangle\,.
\end{eqnarray}
Here the function $F_{N^*}(\vec{q},\vec{p}\,)$ describes the propagation 
of the $NN^*$ configuration followed by the decay $N^*\to\eta N$.

With respect to the spin structure of the amplitude, only three types 
of transitions are possible, namely $S=0\to S'=1$, $S=1\to S'=0$, and 
$S=1\to S'=1$. The transition $S=0\to S'=0$ is excluded since it does not
conserve parity. It is easily seen that the operators $\Sigma_1$ through
$\Sigma_4$ contribute to all three types of transitions. The operators
$\Sigma_5$ and $\Sigma_6$, being antisymmetric under exchange of
$\vec{\sigma}_1$ and $\vec{\sigma}_2$, appear only in channels with a
change of spin, i.e.\ $S=0\to S'=1$ and $S=1\to S'=0$, whereas
$\Sigma_7$ and $\Sigma_8$, which require flipping of the spins 
of both nucleons, contribute only to $S=1\to S'=1$. 

At low energies, the channel $S=1\to S'=0$ is strongly favored in the
$pp\to\eta pp$ reaction since it allows the final nucleons to be left
in a relative $S$ wave whose contribution is further increased by
FSI. For the reaction $np\to\eta np$ both spin changing transitions
are possible near threshold. Apart from a trivial isospin factor 3
appearing for $T=0$, their relative importance depends crucially on
the interference between $\pi$ and $\rho$
exchange~\cite{Wilkin01}. The transition $S=1\to S'=1$, which requires
either the nucleons to be in a relative $P$ state or the $\eta$ meson
to be emitted at least in a $P$ state is suppressed at low energies. 

After these more formal considerations we will now turn to the outline
of the $\eta$ production model used in the present work. The $\eta$
production mechanism via $N^*$ excitation through meson exchange was
considered in a variety of previous
studies~\cite{Wilkin01,Gedal98,Pena01,Nak02,Baru03}. In the present
work, we have adopted this approach including only $\pi$, $\rho$, and
$\eta$ mesons whose contributions was investigated in greater detail
than those of $\omega$, $\sigma$ and other mesons. The corresponding
Lagrangians are taken in the customary form 
\begin{eqnarray}\label{45}
{\cal L}_{\pi NN} \phantom{^*} &=& i\,g_{\pi NN}\,F_\pi(q^2)\
{\bar u}(\vec{p}\,')
\,\gamma^5\,u(\vec{p}\,)\ \vec{\tau}\cdot\vec{\phi}_\pi\,,\nonumber\\
{\cal L}_{\pi NN^*} &=& g_{\pi NN^*}\,F^*_\pi(\vec{q}^{\,2}_{\pi N})\ 
{\bar u}(\vec{p}\,')
\,u(\vec{p}\,)\ \vec{\tau}\cdot\vec{\phi}_\pi\,,\nonumber\\
{\cal L}_{\eta NN} \phantom{^*} &=& i\,g_{\eta NN}\,F_\eta(q^2)\ 
{\bar u}(\vec{p}\,')
\,\gamma^5\, u(\vec{p}\,)\,\phi_\eta\,,\\
{\cal L}_{\eta NN^*} &=& g_{\eta NN^*}\,F^*_\eta(\vec{q}^{\,2}_{\eta N})\ 
{\bar u}(\vec{p}\,')
\,u(\vec{p}\,)\,\phi_\eta\,,\nonumber\\
{\cal L}_{\rho NN} \phantom{^*} &=& \,F_\rho(q^2){\bar u}(\vec{p}\,')\left\{
g_V\gamma^\mu +i
\frac{g_T}{2M_N}\sigma^{\mu\nu}q_\nu\right\}
u(\vec{p}\,)\ \varepsilon_\nu\  \vec{\tau}\cdot\vec{\phi}_\rho\,,\nonumber\\
{\cal L}_{\rho NN^*} &=& i\,F_\rho(q^2)\ {\bar u}(\vec{p}\,')
\frac{g^*_T}{M_{N^*}-M_N}\sigma^{\mu\nu}q_\nu\, u(\vec{p}\,)\ 
\varepsilon_\nu\ \vec{\tau}\cdot\vec{\phi}_\rho\,,\nonumber
\end{eqnarray}
where $q=p-p'$ denotes the four momentum of the exchanged 
meson and $\vec q_{\alpha
N}=(M_N \vec q-m_\alpha \vec p\,)/(M_N+m_\alpha)$ the relative meson
nucleon three momentum with $\alpha\in\{\pi,\eta\}$. Further, each
meson nucleon vertex $\alpha NN$ with $\alpha \in\{\pi,\eta,\rho \}$
contains a form factor of monopole type 
\begin{equation}\label{46}
F_\alpha(q^2)=
\frac{\Lambda_\alpha^2-m_\alpha^2}{\Lambda_\alpha^2-q^2}\,.
\end{equation}
The corresponding $\alpha NN^*$ transition vertices with 
$\alpha \in\{\pi,\eta \}$ are regularized by form factors of the form
\begin{equation}\label{47}
F^*_\alpha(\vec{q}^{\,2}_{\alpha N})=
\frac{\Lambda^{* 2}_\alpha}{\Lambda^{* 2}_\alpha+\vec{q}\,^2_{\alpha N}}\,.
\end{equation}
For the $\rho NN^*$ vertex the same form factor as for $\rho NN$ is used.
With this meson exchange model the functions $M_j$ appearing in
(\ref{25}) can be written as a sum over the participating mesons 
\begin{equation}\label{50}
M_j=M_j^{(\pi)}+M_j^{(\rho)}+M_j^{(\eta)}\,.
\end{equation}
Detailed expressions for $M_j^{(\alpha)}$ $\alpha\in\{\pi,\eta,\rho\}$ 
are listed in the appendix. 

Following the work of \cite{Wilkin01}, the main calculation is based on 
the assumption that the $NN\to NN^*$ transition is dominated by 
$\rho$ exchange. The contribution of $\pi$ meson exchange to the amplitude is 
about three times smaller. The $\eta$ meson with an $\eta NN$ coupling
$g_{\eta NN}^2/4\pi = 0.4$ from~\cite{Tiat94} provides only a tiny
correction at the level of about 3~\%. The $\rho NN^*$ coupling constant 
$g_T^*$ was treated as a free parameter. It was adjusted in such a way that 
the total cross section in both channels, $pp$ and $np$, is normalized to 
the corresponding data with the same normalization factor $C_n < 1$. 
This factor takes effectively into account the damping of the initial
beam via inelastic collisions other than $NN\to\eta NN$.

The parameters appearing in (\ref{45}) to (\ref{47}) are listed in
Tables \ref{table1} and \ref{tab2}. Those for the $\alpha NN$ couplings, 
used also in~\cite{Wilkin01} are given in~\cite{Mach01}, except for
$g_{\eta NN}$ which was taken from~\cite{Tiat94}. The $\pi NN^*$ and
$\eta NN^*$ parameters are taken from~\cite{BeTa91} and \cite{FiAr03}
as is explained below. With this set of parameters we obtain as
normalization factor $C_n = 0.18$, which is appreciably smaller than
the value $C_n \approx 0.55$ of~\cite{Wilkin01} and also less than the
values 0.23 and 0.2 obtained respectively in~\cite{Nak02} and
\cite{Pena02}. In view of some obvious limitations adopted here for
the production potential, e.g., neglect of direct emission of an
$\eta$ meson by one of the nucleons as well as the contribution of
other mesons, we do not put too much significance on this quite small
value for the damping factor $C_n$. 

Clearly, the present model is only one of possible ways to parametrize
the $NN\to NN^*$ transition amplitude. As was discussed
in~\cite{Wilkin01}, the isovector exchange with a dominance of $\rho$
over $\pi$ exchange is consistent with the observed ratio of about 6.5
of the $np\to\eta np$ to $pp\to\eta pp$ cross section \cite{Calen98}
as well as with the characteristic $(a-\cos^2\theta_\eta)$-dependence
of the $\eta$ angular distribution \cite{Calen99}. For this model the
transition with $T=0$ strongly dominates over $T=1$ and, among others,
results in a large probability of deuteron formation~\cite{Wilkin01}
in agreement with experimental results~\cite{Calen98}. However, as has
been shown in~\cite{Nak02} and \cite{Baru03}, more sophisticated
treatments of the initial state interaction as well as inclusion of
genuine two-body meson production mechanisms favour $\pi$
dominance. Therefore, in order to broaden our theoretical basis, we
consider in addition the second type of $NN\to NN^*$ potential, where
the $\rho$ meson is neglected. Another purpose of using such an
extreme approximation is to see to which extent the FSI effects depend
on the details of the $NN\to NN^*$ transition. One might anticipate
the result that due to the short-range nature of the $NN\to NN^*$
interaction such a dependence can not be very significant. However, in
view of the basic theoretical uncertainties with respect to the
production mechanism, this question has to be investigated
quantitatively. Indeed, as will be shown below, the sensitivity of FSI
effects to the variation of the relative strength of $\pi$ to $\rho$
exchanges in the production potential, although not very dramatic,
results in a variation of FSI effects of about 30~\%. 


\section{Results and discussion}\label{sect2}

Our method of solving the three-body equations is based on separable
representations of the driving two-body forces. For the $\eta N$
interaction we use the isobar model of \cite{BeTa91}, where the
resonance parametrization of the partial wave amplitudes is
fitted to the $\pi N\to \pi N$, $\pi N\to \pi\pi N$, and $\pi^- p\to
\eta n$ channels. In particular in the $S_{11}$ channel only the
$S_{11}(1535)$ resonance was taken into account. Apart from the
original set of $S_{11}$ parameters presented in \cite{BeTa91}, which
gives for the $\eta N$ scattering length $a_{\eta N} =
(0.25+i0.16)$~fm, we considered another parametrization listed in
Table~\ref{tab2}, which gives a value twice as large, namely $a_{\eta
N} = (0.52+i0.33)$~fm, and at the same time allows one to describe
well the above-listed reactions. More details are given
in~\cite{FiArPR}. These two sets of parameters are used to study the
sensitivity of the result to the $\eta N$ interaction model. 

The $NN$ interaction is included in $S$, $P$ and $D$ waves in both
isospin states, namely $^1S_0$, $^3P_J$ ($J=0,1,2$), and $^1D_2$ waves
for $T=1$ and $^3S_1$, $^1P_1$, and $^3D_J$ ($J=1,2,3$) for $T=0$. The
corresponding off-shell amplitudes were obtained using the separable
representation of the Paris potential from~\cite{Haid84}. The
remaining higher partial waves of the $NN$ subsystem were taken as plane
waves. 

The Faddeev equations for the $\eta NN$ system were solved only for
the states of lowest angular momentum where a relative two-particle
$S$ state, either $(NN)$ or $(\eta N)$, is coupled with an $S$ state
of the third particle relative to this pair, resulting in the
partitions $\eta+(NN)$ or $N+(\eta N)$, respectively. 
As the explicit calculation shows, in the energy region of
current interest, $Q < 60$ MeV, these $S$ waves 
are almost totally overlapping. The other
partial waves were treated perturbatively up to first order in the
$\eta N$ and $NN$ scattering amplitudes. 

\begin{figure}
\includegraphics[width=30pc]{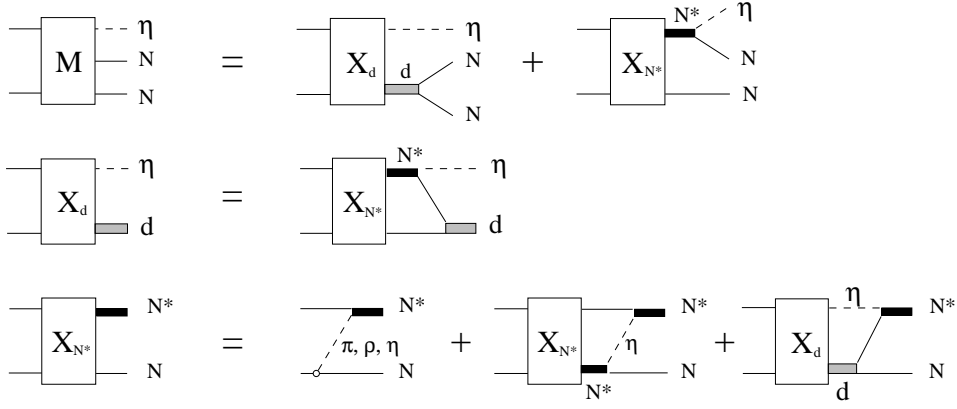}
\caption{\label{fig1}
Diagrammatic representation of the three-body integral equations  
for the reaction $NN\to\eta NN$.} 
\end{figure}

The separable representation of the two-body $t$ matrices allows one
to reduce the original Faddeev equations to a set of two integral
equations in one variable only, presented schematically in
Fig.~\ref{fig1}. Their structure is identical to the usual
Lippmann-Schwinger equation for the two coupled channels $(N+N^*)$ and
$(\eta + d)$, where the notation ``$d$'' is used for the interacting
$NN$ subsystem. The reaction amplitude $M$ is then expressed in terms
of the auxiliary quasi-two-body amplitudes $X_d$ and $X_{N^*}$ as is
presented in the same figure. With respect to the application of
three-body scattering theory to reactions with three free particles in
the final state we refer to~\cite{ScZi74}. Some technical points
concerning the $\eta NN$ problem are described in \cite{FiAr02}. 

\begin{figure}
\includegraphics[width=35pc]{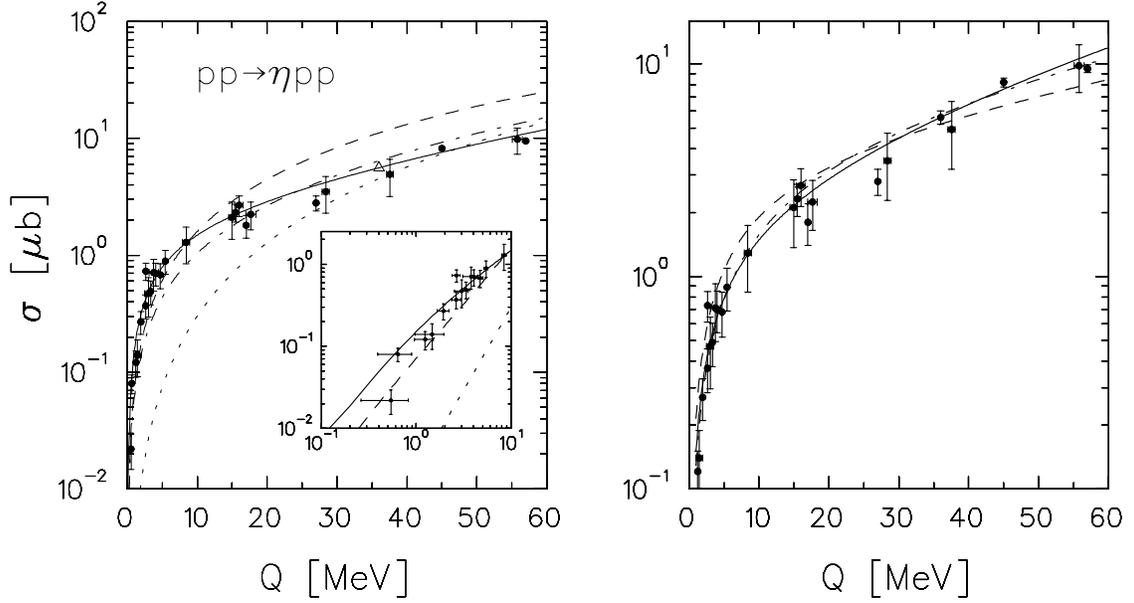}
\caption{\label{fig2}Left panel:
Total cross section for $pp\to\eta pp$. Notation of curves: 
dotted: impulse approximation;
dashed: rescattering in two-body subsystems only; 
dash-dotted: rescattering in $NN$ subsystem only; 
full: three-body calculation. 
Data from compilation~\protect\cite{Mosk03}. Open triangle: 
normalization point $(Q,\sigma)=(36,5.6)$ from 
\cite{Calen96}. Inset shows near-threshold region. 
Right panel: Dependence of three-body result on model ingredients. 
Notation of curves: solid identical to the one in the left panel; dashed:
without $\rho$ exchange in production potential; dash-dotted:
calculation with $S_{11}(1535)$ parametrization from \protect\cite{BeTa91}
with weaker $\eta N$ interaction, $a_{\eta N}=(0.25+i0.16)$
fm, and $\pi$ dominated $NN\to NN^*$ potential.}
\end{figure}

\begin{figure}
\includegraphics[width=18pc]{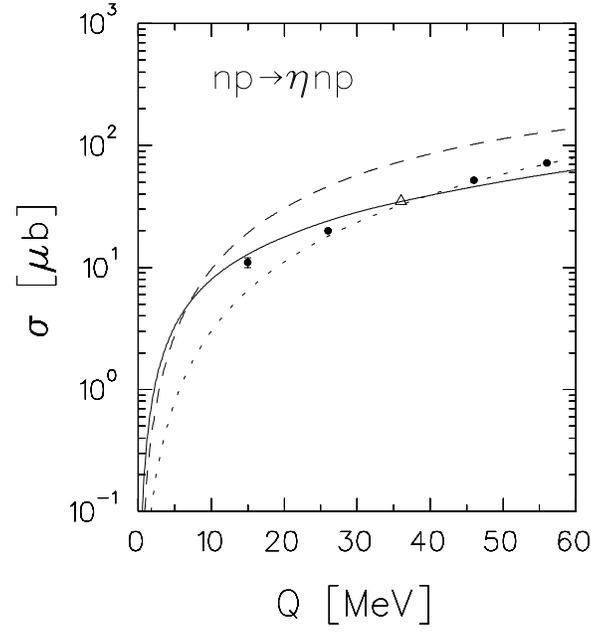}
\caption{\label{fig3}
Same as the left panel of Fig.~\protect\ref{fig2} but for 
$np\to\eta np$. Data from \protect\cite{Calen98}.
} 
\end{figure}

We begin the discussion of our results with the total cross section
shown in Figs.~\ref{fig2} and \ref{fig3} for $pp\to\eta pp$ and
$np\to\eta np$, respectively. As already noted, the three-body result
is normalized to the data at one point, namely at $Q = 36$ MeV to
$\sigma=5.6~\mu$b using PROMISE-WASA data~\cite{Calen96} for $pp$ and
to $\sigma=35~\mu$b with CELSIUS data~\cite{Calen98} for $np$. Of
course, the ratio between the different curves is fixed by the
model. Therefore, apart from the fact that the same normalization
factor is used for both reactions, the only nontrivial result in
Figs.~\ref{fig2} and \ref{fig3} is the energy dependence of the cross
sections. For obvious reasons, detailed experimental data are
available for the $pp\to\eta pp$ reaction only, which, however, is
much more difficult for a theoretical study of the near threshold
region because of the Coulomb interaction between the protons. In the
present work we have not included the Coulomb force, which fact, of
course, hampers to some extent a quantitative comparison with the
data. Nevertheless, the results shown in Figs.~\ref{fig2} and
\ref{fig3} allows us to draw the following conclusions:

The three-body model reproduces quite well the energy dependence of
the experimental cross section in the $pp$ channel. The deviation very
close to threshold must primarily be ascribed to the neglect of
Coulomb repulsion. As for the $np$ channel, the agreement with the
data of \cite{Calen98} appears to be less satisfactory. The theory
exhibits a more flat form of the cross section. Clearly, for a
meaningful conclusion about the quality of the present model a greater
amount of data is needed, especially in the vicinity of the threshold. 

The first-order calculation, where the FSI is reduced to
only $\eta N$ and $NN$ rescatterings, does not reproduce the form of
the $pp\to\eta pp$ cross section as predicted by the three-body
treatment and also observed in the experiment. As was noted in the
introduction, it is expected that the leading terms associated with
$\eta N$ and $NN$ rescatterings do not constitute a reliable
approximation of the whole multiple scattering series, due to the slow
convergence of this series in the
vicinity of the virtual $\eta NN$ pole. The strong difference between
the first order and the exact three-body calculation even at $Q = 60$
MeV demonstrates the crucial importance of higher order terms. We
consider this result as a direct evidence for the fact that a
three-body approach is definitely needed for a correct treatment of
the $\eta NN$ interaction. Thus, the first-order approximation,
although seemingly reasonable in view of the weak $\eta N$ interaction
as compared to the $NN$ one, is in general not relevant.  

\begin{figure}
\includegraphics[width=35pc]{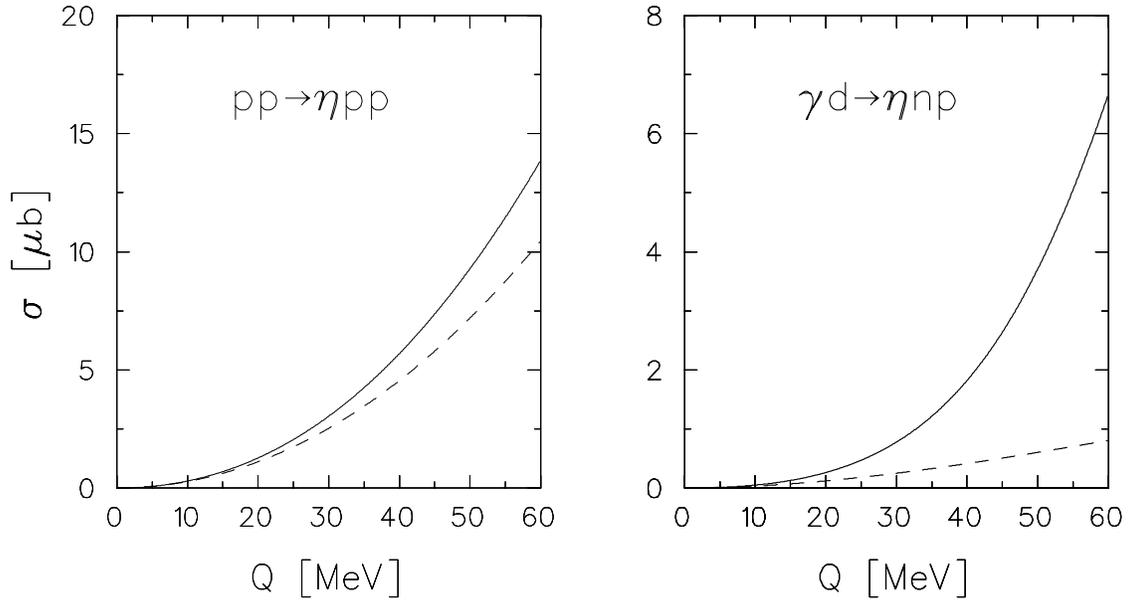}
\caption{\label{fig4}
Total cross sections for $pp\to\eta pp$ (left panel) and 
$\gamma d\to\eta np$ (right panel) calculated in the impulse
approximation. Dashed lines include only $^1S_0s$   
wave in the final state.}  
\end{figure}

The question of a correct treatment of FSI in $\eta$ production is 
closely related to the relative fraction of $S$ waves in the final 
state. As was mentioned above, in accordance with the short range 
hypothesis for the $\eta$ production mechanism, the $S$ waves in the 
$\eta NN$ system are expected to provide the major part of the cross 
section in the threshold region, thus making the FSI effect to be 
of particular importance in this reaction. In this context it is 
instructive to compare the relative contribution of the $S$ waves to 
the total cross sections of $\gamma d\to\eta np$ and $pp\to\eta pp$. 
The results of the IA for both processes are presented in Fig.~\ref{fig4}, 
where the $^1S_0s$ contribution~\cite{fnote} is shown separately. 
Because of the relatively long range 
nature of the mechanism for $\eta$ photoproduction on the deuteron, 
the contribution of higher partial waves in the final state increases 
rapidly and already at the energy $Q = 60$ MeV the $S$ wave provides 
only a 10 $\%$ fraction of the total cross section. In contrast to 
this, the $^1S_0s$ configuration strongly dominates the reaction 
$pp\to\eta pp$ in the whole energy region, so that even at $Q = 60$ 
MeV it produces about 77 \% of the cross section. In conclusion, the 
role of the final state interaction depends not only on the phase 
space available for the final particles (which is the same for both 
reactions) but much more crucially on the range of the primary 
$NN\to NN^*$ transition interaction. 

On the right panel of Fig.~\ref{fig2} the dependence of the
results on the model ingredients is exhibited. 
Firstly, we would like to note a visible
sensitivity of the cross section to the choice of which meson exchange
governs the production potential. Namely, $\pi$ dominance 
tends to make the form of the total cross section
more convex as compared to the full curve where $\rho$ exchange prevails,  
whereas the IA gives in both cases practically indistinguishable
results except for the normalization factor $C_n$. 
Using a weaker $\eta N$ interaction from
\cite{BeTa91}, we obtain, as expected, a weaker influence of FSI as is shown
by the dash-dotted line in the right panel of Fig.~\ref{fig2}.  

\begin{figure}
\includegraphics[width=35pc]{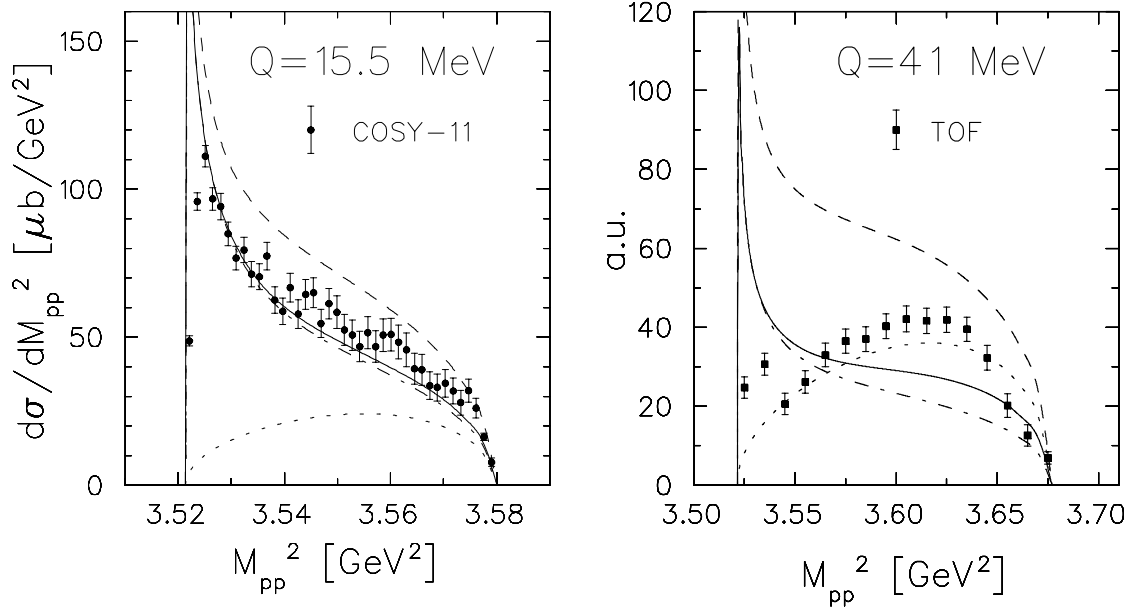}
\caption{\label{fig5}
Distribution of squared $pp$ invariant mass in $pp\to\eta
pp$. Notation 
of curves as in the left panel of Fig.~\protect\ref{fig2}. The dash-dotted
curve is obtained with a pure $^1S_0s$ final state. Data
at $Q=15.5$ MeV are from \protect\cite{Mosk03} and at $Q=41$ MeV
from \protect\cite{TOF}. 
} 
\end{figure}

In Fig.~\ref{fig5} we compare our calculation for the $pp$ invariant mass 
distribution with the COSY-11~\cite{Mosk03} and TOF~\cite{TOF} data.  
At the excess energy $Q = 15.5$ MeV 
the three-body result was normalized to the experimental
integrated cross section $\sigma = 3.24$ $\mu$b of~\cite{Mosk03}. 
The distribution at $Q = 41$ MeV is given in arbitrary units and 
the theoretical full curve was normalized to the area enclosed by the data
points. For a lower excess energy $Q=15.5$ MeV one 
finds a reasonable agreement of the three-body calculation with the data,
although one notes a systematic slight underestimation of the 
experimental cross section in the region 
above $M_{pp}^2=3.54$~(GeV)$^2$. At least some 
part of the deviation stems again from an overestimation of the $pp$ 
interaction effect at the lower end of the spectrum due to the neglect 
of the Coulomb force. In view of the overall normalization, inclusion 
of the Coulomb repulsion would reduce the $^1S_0$ peak near $M_{pp}=2M_N$, 
and would enhance the other part of the distribution. 

However, at the higher excess energy $Q=41$ MeV the 
shape of the theoretical distribution is in strong disagreement with the data. 
In particular, the broad maximum around $M^2_{pp}= 3.62$~GeV$^2$ is not
explained. Of course, in order to reach at a definite conclusion, the Coulomb
interaction  must be included in addition, but it is very unlikely that 
this would have such a strong effect as to explain the shape. Moreover, 
our results are at variance with those 
of~\cite{Nak03} where an appreciable contribution
of higher angular momentum states in the $pp$ subsystem is predicted, 
whereas our calculation does not indicate a sizeable influence 
from $P$ and $D$ partial $NN$ waves. The corresponding contributions
can be estimated by comparison of the full and the dash-dotted curves 
in Fig.~\ref{fig5}. In the latter calculation only the lowest $^1S_0s$ 
configuration in the $\eta NN$ final state was included. 

\begin{figure}
\includegraphics[width=18pc]{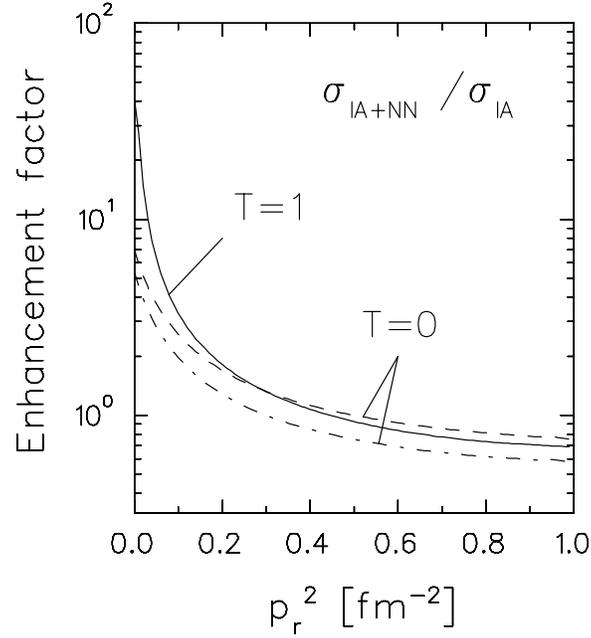}
\caption{\label{fig6}
Enhancement factors associated with the $NN$ interaction in $T=0$ and $T=1$
channels as function of the squared relative $NN$ momentum. 
The dashed  and dash-dotted curves are obtained with and without 
inclusion of the $^3D_1\to {^3S_1}$ transition, respectively.  
}  
\end{figure}

On the other hand, it is worth to underline the importance of the $NN$ tensor 
force, which increases quite appreciably the transition to the $^3S_1$ 
state from an intermediate $^3D_1$ after $\eta$ production. The relevance 
of this intermediate $^3D_1$ state is related to the strong momentum
transfer accompanying $\eta$ production and emphasizing the high momentum
components of the $NN$ wave function. One of the consequences of this
fact is that the enhancement by the $NN$ interaction for the $T=0$ 
transition is even slightly stronger than that for $T=1$. This result is 
at variance with~\cite{Wilkin01}, where a ratio of the enhancement 
factors of $T=1$ to $T=0$ of 1.85 at relative $NN$ momenta $p_r > 0.5$ 
fm$^{-1}$ was found. According to our results in Fig.~\ref{fig6}, this 
ratio is reduced from 1.35 to only 0.98 at $p_r = 0.5$ fm$^{-1}$  if the 
transition $^3D_1\to\!^3S_1$ is included. 

\begin{figure}
\includegraphics[width=35pc]{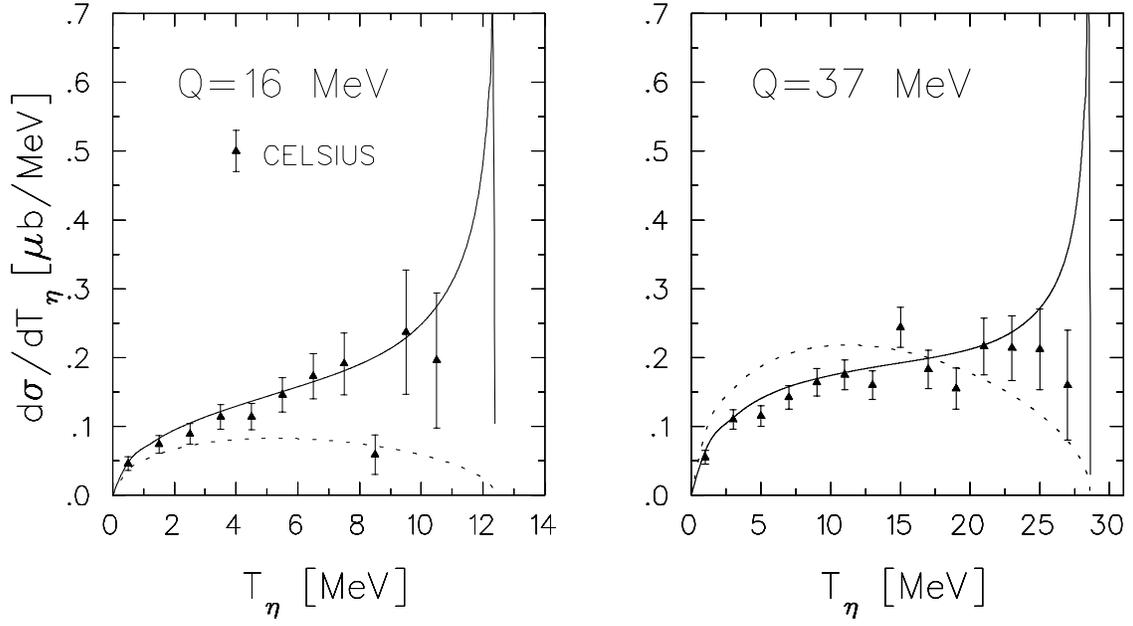}
\caption{\label{fig7}
$\eta$ meson spectrum for $pp\to\eta pp$ as a function of $\eta$
kinetic energy $T_\eta$. Solid and dotted
curves show respectively the three-body model and IA predictions. 
Data from \protect\cite{Calen99}.
} 
\end{figure}

In Fig.~\ref{fig7} the calculated distribution with respect to
the $\eta$ kinetic energy $T_\eta$ in $pp\to\eta pp$ is plotted together 
with the CELSIUS data \cite{Calen99}. One readily notes quite a good
agreement except for a very small region
close to the maximal kinetic energy. We do
not consider this discrepancy as a serious defect because, as was
noted in~\cite{Calen99}, some fraction of the events must have been lost. 
Furthermore, we would like to point out that the data of the $pp$ mass
distribution at $Q=41$~MeV in the right panel of Fig.~\ref{fig5}
appears to be in conflict with the data at $Q=37$~MeV in the right
panel of Fig.~\ref{fig7}. The latter increases monotonically
with meson energy except very close to the maximal kinetic energy
$T_\eta$, whereas the TOF results for $d\sigma/dM^2_{pp}$ in
Fig.~\ref{fig5} exhibit a pronounced broad maximum at intermediate $pp$
masses. In view of the simple relation 
\begin{equation}
\frac{d\sigma}{dT_\eta} = -2W\frac{d\sigma}{dM^2_{pp}} \,,
\end{equation}
one would expect on the other hand a similar behaviour for both
distributions, because the small difference in the energies 
can hardly result in a very different form of the spectrum. 
Thus new experimental data for the $T_\eta$ and/or
$M_{pp}^2$ distributions in the region $Q = 37 - 41$ MeV are needed in
order to resolve this discrepancy. 

\begin{figure}
\includegraphics[width=35pc]{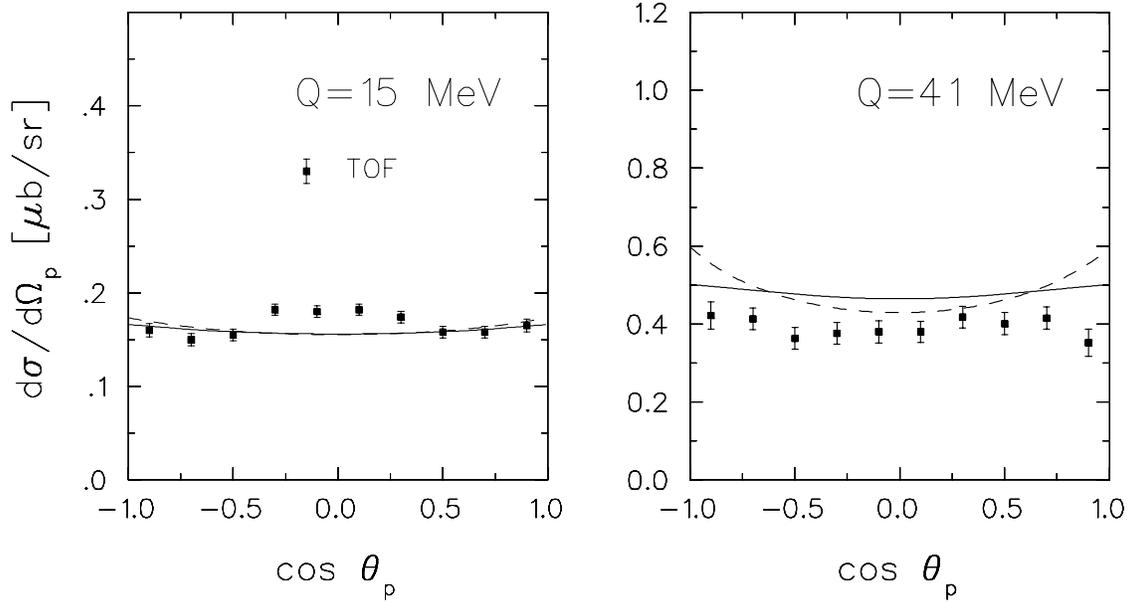}
\caption{\label{fig8}
Angular distribution of protons in $pp\to\eta pp$. In the dashed 
curves only $\pi$ and $\eta$ meson exchange is included in the production
potential. Data from \protect\cite{TOF}.
} 
\end{figure}

Finally, we present in Fig.~\ref{fig8} the angular distribution of an
emitted proton for two excess energies. The dominance of $\rho$ exchange 
in the production mechanism leads to a pronounced isotropy of the 
cross sections. This becomes apparent by comparison with the
angular distributions using a pure $\pi$ exchange.
They exhibit a much stronger angular dependence with a more pronounced
peaking at forward and backward directions, which is consistent with a
more periferal character of the $\pi$ exchange mechanism. 
In general, the proton angular distribution with a weak
$\cos^2\theta_p$ dependence is consistent with the almost 
isotropic form of the experimental results of \cite{TOF}, although the
observed magnitude at $Q=41$ MeV is overestimated by about 20 
\%. Furthermore, one should keep in mind the pronounced disagreement
between theory and the TOF data for the $M^2_{pp}$ spectrum at the same
energy (see Fig.~\ref{fig5}). With respect to
the $\eta$ angular distribution, we did not analyze it here in view
of its obvious sensitivity to higher $\eta N$ resonances, predominantly 
$D_{13}(1520)$, which were omitted in the present calculation.

\section{Conclusions}\label{sect4}
In the present paper we have investigated $\eta$ production in $NN$
collisions in the near threshold region. The main attention was
focussed on the understanding of the role of FSI, which is expected to
strongly influence the reaction dynamics near threshold. We
believe that in conjunction with previous
work~\cite{Wilkin01,Gedal98,Pena01,Nak02,Baru03} the present study was
necessary in order to establish a unified picture of this reaction. In
particular, it can be quite useful for the planning and analysis of
future experiments. From our results we draw the following conclusions:

\noindent
(i) The three-body treatment of the $\eta NN$ interaction provides a
satisfactory, quantitative explanation of the energy dependence of
the total cross section for $pp\to\eta pp$. However, the dependence of
the results on the details of the $NN\to NN^*$ transition introduces
some uncertainty. In particular, assuming a pion exchange dominance in
the production potential leads to stronger energy dependence of the
cross section and requires, therefore, a weaker $\eta N$ interaction
in order to bring the theory into agreement with the data. 

\noindent
(ii) There is a discrepancy between our results and the data
of~\cite{TOF} for the $M^2_{pp}$ distribution at a c.m.\
excess energy $Q=41$ MeV. Namely, the broad peak of the data is not
described by the theory. Comparing to other theoretical work, we do
not reproduce the result of~\cite{Nak03}, where this broad structure
is explained by the contribution of
higher partial waves in the $NN$ subsystem. On the contrary, according
to our findings the contribution of $P$  and $D$ waves in the final
$pp$ subsystem plays a minor role only, at least in the region up to
$Q = 60$ MeV considered in the present work.  

\noindent
(iii) Even close to threshold, where the final $NN$ system is
predominantly in an $S$ wave state, the tensor $NN$ force in the $T =
0$ channel must be taken into account. As we have shown, the
intermediate transition to a $^3D_1$ configuration with subsequent
rescattering into the $^3S_1$ state enhances sizeably the $NN$
interaction effect in the triplet state. 

Before concluding, we would like to note that the sensitivity of the FSI
effect to the $NN\to NN^*$ transition mentioned in (i) indicates 
that in principle the short range part of the final state wave
function must be calculated more thoroughly than is done
here. Obviously, the most straightforward way would be a three-body
treatment of the whole $NN\to\eta NN$ process without restriction to
the final state. However it seems to be very difficult in view of the great
amount of channels which have to be included in such an approach, not
to mention the pure theoretical ambiguity in treating the $NN$ system within
the three-body formalism. On the other hand, one can expect that short range
mechanisms may be included perturbatively. For example, the $\pi NN$
configurations could be considered in first order only (apart from the driving
meson exchange potential) by including the additional mechanism where a pion,
created in an $NN$ collision, produces in turn an $\eta$ via rescattering on
one of the nucleons. Clearly, such an approach, which would allow us
to adopt the same FSI concept, is practically much easier to manage
than a rigorous solution of the three-body coupled channel model.

\acknowledgments
This work was supported by the Deutsche Forschungsgemeinschaft (SFB 443). 

\renewcommand{\theequation}{A\arabic{equation}}
\setcounter{equation}{0}
\section*{Appendix A}
\label{appA}

In this appendix we give the explicit expressions for the amplitudes
$M^{(\alpha)}_j$, $\alpha\in\{\pi,\eta,\rho \}$ for the
transition $NN\to NN^*$ determining the transition matrix element
$T_{NN^*}(\vec{p},\vec{p}\,')$. The latter was defined in the c.m.\
system with $\vec{p}$ and $\vec{p}\,'$ denoting the momenta of the
initial and final states, respectively. In the following expressions
we have introduced the notations:
\begin{equation}
\begin{array}{lll}
E=\sqrt{p^2+M_N^2}=\frac{W}{2}\,,\quad &  E'=\sqrt{p^{\prime\,2}+M_N^2}\,, 
& E^*=\sqrt{p^{\prime\,2}+M_{N^*}^{2}}\,,        \cr
\epsilon = E+M_N\,, & \epsilon' = E'+M_N\,, & \epsilon^* = E^*+M_{N^*}\,,\cr
P_\mu = (E+E',\,\vec{p}+\vec{p}\,')\,, & 
P^*_\mu = (E+E^*,\,-\vec{p}-\vec{p}\,')\,,\quad  &  
\Delta M^*=M_{N^*}-M_N\,,\cr
\vec q^{\,\prime (*)}= \frac{\vec p}{\epsilon}
- \frac{\vec p^{\,\prime}}{\epsilon^{\prime (*)}}\,, &
\vec r^{\,\prime (*)}= \frac{\vec p}{\epsilon}
+ \frac{\vec p^{\,\prime}}{\epsilon^{\prime (*)}}\,. & \cr
\end{array}
\end{equation}

The expressions for $\rho$ exchange are obtained by using
the on-shell Gordon decomposition of the nucleon current 
\begin{equation}\label{a10}
\bar{u}(\vec{p}\,')\,\gamma_\mu\, u(\vec{p}\,)=\frac{1}{2M_N}\,
\bar{u}(\vec{p}\,')\left[(p_\mu+p'_\mu) - i\sigma_{\mu\nu}(p_\nu-p'_\nu)\right]
u(\vec{p}\,)
\end{equation}
and the analogous relation for the $\rho NN^*$ coupling. In the
present paper the invariant energy $\omega_{N^*}$ of the $N^*$ or
$\eta N$ subsystem was determined according to the spectator on-shell
scheme 
\begin{equation}
\omega_{N^*}=
W-M_N-\frac{p^2}{2}(\frac{1}{M_N}+\frac{1}{M_N+m_\eta})\,,
\end{equation}
which is consistent with the nonrelativistic three-body treatment of
the $\eta NN$ system. Furthermore, this choice of $\omega_{N^*}$
justifies the use of the Gordon decomposition for the tensor part in the
$\rho NN$ coupling since in this case the nucleon remains on-shell
after emission of a $\rho$ meson. The only approximation, whose validity
is not justified is the Gordon relation for the $\rho NN^*$
coupling. We believe, however, that, since the coupling constant
$g^*_{T}$ is treated here as a free parameter, it is a reasonable
approximation.  

Using the vertices (\ref{45}) one finds the following contributions 
from $\pi$, $\eta$, and $\rho$ meson exchange:

\noindent
(i) $\pi$ meson exchange: 
\begin{equation}\label{a15}
M^{(\pi)}_j= \frac{\epsilon}{2M_N}\sqrt{\frac{\epsilon'\epsilon^*}
{4M_NM_{N^*}}}\ g_{\pi NN}\,g_{\pi NN^*}\frac{F_\pi(q^2)F^*_\pi
(\vec q_{\pi N}^{\,2})}{q^2-m^2_\pi} \,F_j\,,
\end{equation}
where 
\begin{eqnarray}\label{a20}
F_{1(2)}&=&\pm\frac{1}{\epsilon^{(\prime)}}\Big(1-\frac{\vec{p}\cdot\vec{p}\,'}
{\epsilon\epsilon^*}\Big)\,,\\
F_{3(4)}&=0\,,\\
F_{6(5)}&=&\mp\frac{\vec q\cdot\vec p^{\,(\prime)}
}{\epsilon\epsilon^*}\,, \\
F_{7(8)}&=&\pm\frac{1}{\epsilon\epsilon^*\epsilon^{(\prime)}}\,.
\end{eqnarray}

\noindent
(ii) $\eta$ meson exchange: Since the formal expressions are the same
as for $\pi$ exchange except for a replacement of coupling constants
and meson mass, we do not need to repeat them here:
\begin{equation}\label{a25}
M^{(\eta)}_j=M^{(\pi)}_j\Big|_{\pi\to\eta}\,.
\end{equation}

\noindent
(iii) $\rho$ meson exchange: 
\begin{equation}\label{a30}
M^{(\rho)}_j=-\frac{\epsilon}{2M_N}\sqrt{\frac{\epsilon'\epsilon^*}
{4M_NM_{N^*}}}\ \frac{g_T^*\ F_\rho^2(q^2)}
{q^2-m^2_\rho} \,F_j\,,
\end{equation}
where 
\begin{eqnarray}\label{a35}
F_{2(1)}&=&\mp(g_V+g_T)\frac{\vec q^{\,\prime}\cdot\vec p^{\,(\prime)}}
{\epsilon\epsilon^*}\\
F_3&=&\frac{g_V+g_T}{\epsilon}\Big[
-2-\frac{\vec{p}\,'\cdot \vec r^{\,*}}{\epsilon'}
+\Big(1+\frac{\vec{p}\cdot\vec{p}\,'}
{\epsilon\epsilon'}\Big)\frac{P_0^*}{\Delta M^*}
+\frac{\vec r^{\,\prime}\cdot\vec P}{\Delta M^*}
\Big]\nonumber\\
&+&
\frac{g_T}{2M_N\epsilon}
\Big(1-\frac{\vec{p}\cdot\vec{p}\,'}{\epsilon\epsilon'}\Big)
\Big[W+\epsilon'+\frac{p^{\prime\,2}}
{\epsilon^*}-\frac{P^\mu P^*_\mu}{\Delta M^*}
\Big]\,,\\
F_4
&=&\frac{g_V+g_T}{\epsilon^*}\Big[
-1-\frac{\epsilon^*}{\epsilon'}-\frac{\vec{p}\cdot\vec r^{\,\prime}}{\epsilon}
-\Big(1+\frac{\vec{p}\cdot\vec{p}\,'}{\epsilon\epsilon'}\Big)
\frac{P_0^*}{\Delta M^*}-\frac{\vec r^{\,\prime}\cdot\vec P}{\Delta M^*}
\Big]\nonumber\\
&+&
\frac{g_T}{2M_N\epsilon^*}
\Big(1-\frac{\vec{p}\cdot\vec{p}\,'}{\epsilon\epsilon'}\Big)
\Big[W+\epsilon+\Delta M^*+\frac{p^2}{\epsilon}
+\frac{P^\mu P^*_\mu}{\Delta M^*}\Big]
\\
F_5
&=&-\frac{g_V+g_T}{\epsilon\epsilon'}
\Big[2\epsilon'+\frac{\vec{p}\cdot\vec{p}\,'}
{\epsilon}\Big(1-\frac{\epsilon'}{\epsilon^*}\Big)
+\frac{2p^{\prime\,2}}{\epsilon^*}
-\frac{2(W+M_N)}{(\Delta M^*)}\,\vec p^{\,\prime}\cdot\vec q^{\,*}\Big]
\nonumber\\
&-& \frac{g_T}{2M_N\epsilon\epsilon'}
\Big[\frac{\vec{p}\cdot\vec{p}\,'}{\epsilon}\Big(W+\epsilon'+\frac{p^{\prime\,2}}
{\epsilon^*}\Big)+
\frac{p^{\prime\,2}}{\epsilon^*}\Big(W+\epsilon'+\Delta M^*+
\frac{p^2}{\epsilon}\Big)
-\vec p^{\,\prime}\cdot\vec q^{\,*}\,\frac{P^\mu P^*_\mu}{\Delta M^*}\Big]
\\
F_6
&=&\frac{g_V+g_T}{\epsilon\epsilon'}
\Big[2\epsilon+\frac{p^2}
{\epsilon}\Big(1+\frac{\epsilon'}{\epsilon^*}\Big)-
\frac{2(W+M_N)}{\Delta M^*}\,\vec p\cdot\vec q^{\,*}\Big]
\nonumber\\
&+&\frac{g_T}{2M_N\epsilon\epsilon'}
\Big[\frac{\vec{p}\cdot\vec{p}\,'}{\epsilon^*}\Big(W+\epsilon+\Delta M^*+
\frac{p^2}{\epsilon}\Big)+
\frac{p^2}{\epsilon}\Big(W+\epsilon'+\frac{p^{\prime\,2}}
{\epsilon^*}\Big)
-\vec p^{\,\prime}\cdot\vec q^{\,*}\,
\frac{P^\mu P^*_\mu}{\Delta M^*}\Big]
\\
F_7
&=&\frac{g_V+g_T}{\epsilon^2\epsilon'}\Big[
1+\frac{\epsilon'}{\epsilon^*}
-\frac{2(W+M_N)}{\Delta M^*}\Big]
+
\frac{g_T}{2M_N\epsilon^2\epsilon'}\Big[W+\epsilon'+
\frac{p^{\prime\,2}}{\epsilon^*}-
\frac{P^\mu P^*_\mu}{\Delta M^*}\Big]
\\
F_8
&=&\frac{g_V+g_T}{\epsilon\epsilon'\epsilon^*}\Big[2+
\frac{2(W+M_N)}{\Delta M^*}\Big]
+\frac{g_T}{2M_N\epsilon\epsilon'\epsilon^*}
\Big[W+\epsilon+
\Delta M^*+\frac{p^2}{\epsilon}+
\frac{P^\mu P^*_\mu}{\Delta M^*}\Big]
\end{eqnarray}



\begin{table}[h]
\renewcommand{\arraystretch}{1.2}
\caption{\label{table1}Coupling constants and form factor
parameters of the $\alpha NN$ vertices of (\protect\ref{45}).} 
\begin{ruledtabular}
\begin{tabular}{ccccccc}
$\frac{g^2_{\pi NN}}{4\pi}$ & $\Lambda_\pi$ [GeV] & 
$\frac{g^2_{\eta NN}}{4\pi}$ & $\Lambda_\eta$ [GeV] & 
$\frac{g_V^2}{4\pi}$ & $\frac{g_T^2}{4\pi}$ & $\Lambda_\rho$ [GeV]  \\ \hline
14.4 & 1.7 & 0.4 & 1.5 & 0.84 & 31.25 & 1.4 \\
\end{tabular}
\end{ruledtabular}
\end{table}

\begin{table}[h]
\renewcommand{\arraystretch}{1.2}
\caption{\label{tab2}
The same as in Table \protect\ref{table1} for the $\alpha NN^*$ vertices.}
\begin{ruledtabular}
\begin{tabular}{ccccc}
$\frac{g_{\pi NN^*}}{\sqrt{3}}$ & $\Lambda^*_\pi$ [GeV] & 
$g_{\eta NN^*}$ & $\Lambda^*_\eta$ [GeV] &
$g_T^*$  \\ \hline
1.45 & 0.404 & 2.00 & 0.695 & 1.7  \\
\end{tabular}
\end{ruledtabular}
\end{table}
\newpage

\vspace{1cm}

\vspace{1cm}

\vspace{1cm}

\vspace{1cm}

\vspace{1cm}

\vspace{1cm}


\begin{thebibliography}{99}
\bibitem{fnote}
We use the notation $^{2s+1}L_jl$ for states of the $(\eta+NN)$ partition.
\end{thebibliography}
\end{document}